\documentclass[aps,prc,preprint,groupedaddress,floatfix]{revtex4-2}

\usepackage{graphicx}
\usepackage{amsmath}

\begin{document}


\title{Low-energy quadrupole collectivity of Sn nuclei
  in self-consistent calculations with a semi-realistic interaction}


\author{Y.~Omura$^1$}
\author{H.~Nakada$^2$}\email[E-mail:\,\,]{nakada@faculty.chiba-u.jp}
\author{K.~Abe$^1$}
\author{M.~Takahashi$^2$}
\affiliation{$^1$ Department~of~Physics, Graduate~School~of~Science~and~Engineering,
 Chiba~University,\\
Yayoi-cho~1-33, Inage, Chiba~263-8522, Japan}
\affiliation{$^2$ Department~of~Physics, Graduate~School~of~Science,
 Chiba~University,\\
Yayoi-cho~1-33, Inage, Chiba~263-8522, Japan}


\date{\today}

\begin{abstract}
  Quadrupole collectivity of the lowest-lying states,
  focusing on $E_x(2^+_1)$ and $B(E2;0^+_1\to 2^+_1)$,
  have been investigated for the $N=50-82$ Sn nuclei
  by applying the self-consistent approaches
  with the semi-realistic interaction M3Y-P6.
  Both $E_x(2^+_1)$ and $B(E2;0^+_1\to 2^+_1)$ are well reproduced
  by the spherical Hartree-Fock-Bogolyubov (HFB) plus
  quasiparticle random-phase approximation (QRPA) calculations in $N\geq 64$,
  without adjustable parameters.
  The measured $B(E2)$ values in the neutron-deficient Sn nuclei cast a puzzle.
  In $54\leq N\leq 62$,
  the spherical HFB\,+\,QRPA calculations give too strong $B(E2)$,
  opposite to the shell-model predictions within the one major shell.
  Via the constrained-HFB (CHFB) calculations,
  it is found that the neutron-deficient Sn nuclei are soft
  against the quadrupole deformation,
  accounting for the limited applicability of the HFB\,+\,QRPA approach
  and possibly giving rise to shape fluctuation.
  In particular, the potential energy curves (PECs) are almost flat
  in the range of $|q_0|\lesssim 200\,\mathrm{fm}^2$ in $^{106-110}$Sn.
  We confirm that the near degeneracy of $n0g_{7/2}$ and $n1d_{5/2}$
  triggers weak quadrupole deformation
  and its balance with the pairing makes PECs flat,
  which is qualitatively consistent with a recent shell model result
  in an extended model space,
  by the calculations shifting the single-particle energy spacing
  and the pairing strength.
  These conclusions are supported
  by the proton-to-neutron ratios of the transition matrix elements
  and the reference values of $B(E2)$ with the angular-momentum projection
  on top of the CHFB solutions.
\end{abstract}


\maketitle



\section{Introduction\label{sec:intro}}

The magic numbers, which manifest the shell structure,
are an important and striking property of atomic nuclei.
The nuclei become relatively stable and usually take spherical shapes
when either the proton number $Z$
or the neutron number $N$ is a magic number.
On the other hand,
experiments using radioactive nuclear beams in recent decades
have disclosed that magic numbers are not so rigorous as once expected.
It has been a hot topic how stiff the known individual magic numbers are
and where we may find new magic numbers,
particularly for nuclei off the $\beta$-stability~\cite{ref:SP08}.

With the magic number $Z=50$,
the Sn nuclei have supplied a typical example of semi-magic nuclei.
Their first excitation energies of the even-$N$ Sn nuclei $E_x(2^+_1)$
are almost constant in $N=50-82$~\cite{ref:NuDat}.
This property looks compatible with the seniority
or the generalized seniority scheme~\cite{ref:Tal93},
considering only the degrees-of-freedom (d.o.f.s) of the valence neutrons,
\textit{i.e.}, the neutrons in the $N=50-82$ shell.
The $N$-dependence of the $B(E2;0^+_1\to 2^+_1)$ is also
in good agreement with the seniority scheme
in the $60\lesssim N\leq 80$ region,
once the effective charge for the neutrons is adjusted.
A coupled-cluster calculation with the nucleonic interactions
from the chiral-effective-field theory
gives similar results~\cite{ref:MSS18}.
The effective charge indicates that,
while the doubly magic core is not fully inert and is polarized at $2^+_1$,
the core polarization effects are stable and renormalizable by a constant value.
However, recent data on $B(E2;0^+_1\to 2^+_1)$ in the neutron-deficient region
show significant enhancement compared with the shell-model prediction
within a single major shell~\cite{ref:Ban05}.
This enhancement has cast a question on the conventional picture
of the structure of the Sn nuclei.

To account for the enhancement of $B(E2)$ within the shell model,
an $N$-dependent effective charge was introduced
in Refs.~\cite{ref:JLF12,ref:Doo14}.
On the other hand, a shell-model calculation in an extended model space
suggested that the neutron-deficient Sn nuclei
are slightly deformed~\cite{ref:Tog18}.
The possibility of quadrupole deformation was also suggested
in Ref.~\cite{ref:Zuk21},
and enhanced collectivity via the cross-shell excitations
was investigated in Ref.~\cite{ref:CCG15,ref:KSMS21}.
To avoid ambiguity in the effective charge,
it is desired to treat the core polarization explicitly.
For this purpose, the self-consistent mean-field (MF) approximation
and the random-phase approximation (RPA) on top of it,
which handles all nucleons,
supply an appropriate platform.
In practice, the quasiparticle-RPA (QRPA) approaches
on top of the Hartree-Fock-Bogolyubov (HFB) solutions
have successfully been applied to describe
the low-energy quadrupole collectivity of spherical nuclei
over a wide range of the mass table~\cite{ref:TEB08,
  ref:CTP12,ref:PM14}.
QRPA calculations for the Sn nuclei using the Skyrme and the Gogny interactions
were reported in Refs.~\cite{ref:TEB08,ref:CTP12,ref:PM14,ref:Cor15,ref:LDP15,
  ref:Jun11},
and a relativistic QRPA calculation using the NL3 Lagrangian was reported
in Ref.~\cite{ref:Ans05}.
Also, the five-dimensional collective Hamiltonian (5DCH)
was derived from the HFB results with the Gogny interaction
and applied to the semi-magic nuclei including Sn~\cite{ref:PM14}.
A result of the quasiparticle-phonon model was found
in Ref.~\cite{ref:IST11}.
These results bring in further complexity.
In contrast to the conventional shell-model results,
most of the results of the self-consistent calculations
overestimate the $B(E2)$ values for the neutron-deficient Sn nuclei\cite{
  ref:CTP12,ref:PM14,ref:Nak11},
though the degree significantly depends on the adopted interaction
and the theoretical methods.
It should be noticed that
the good description of the low-energy quadrupole collectivity
does not guarantee success in a specific region,
particularly where the collectivity is strongly influenced
by the shell structure.
Since it was suggested that the $B(E2)$ values could be sensitive
to the shell structure~\cite{ref:Nak11},
the application of the semi-realistic
M3Y-P6 interaction~\cite{ref:Nak13,ref:Nak20},
which gives shell structures compatible
with a large body of the data~\cite{ref:NS14},
will be of interest.
The M3Y-type semi-realistic interaction has an advantage
as it is almost free from unphysical instabilities
against excitations~\cite{ref:DPN21}.

We investigate low-energy quadrupole collectivity of the $^{100-132}$Sn nuclei,
$E_x(2^+_1)$ and $B(E2;0^+_1\to 2^+_1)$ to be precise,
from a microscopic standpoint
using the HFB and QRPA calculations with the semi-realistic interaction
M3Y-P6.
Special interest is in the $B(E2)$ in the neutron-deficient region,
for which accumulated experimental data provided a puzzle.
The numerical setup is presented in Sec.~\ref{sec:setup}.
In Sec.~\ref{sec:QRPA}, we show the results of the QRPA
on top of the spherical HFB solutions.
As weak quadrupole deformation was suggested in the neutron-deficient region,
we have implemented the constrained-HFB (CHFB) calculation
under the axial symmetry,
and examined the possibility of deformation or sphericity,
as discussed in Sec.~\ref{sec:def}.
Summary is given in Sec.~\ref{sec:summary}.

\section{Numerical setup\label{sec:setup}}

We apply the following Hamiltonian in this paper,
\begin{equation} H = K + V_N + V_C - H_\mathrm{c.m.}\,,
\label{eq:Hamil}\end{equation}
where $K = \sum_i {\mathbf{p}_i^2}/(2M)$ with $M=(M_p+M_n)/2$
is the kinetic energy,
$V_N$ is the nucleonic interaction,
$V_C$ is the Coulomb interaction among protons,
and $H_\mathrm{c.m.} = \mathbf{P}^2/(2AM)$
with $\mathbf{P}=\sum_i \mathbf{p}_i$ and $A=Z+N$
is the center-of-mass (c.m.) Hamiltonian.
The subscript $i$ is the index of the constituent nucleons.
For $V_N$, the M3Y-P6 interaction is employed~\cite{ref:Nak13},
which has an origin in a $G$-matrix~\cite{ref:M3Y,ref:M3Y-P},
though several channels have been modified
phenomenologically~\cite{ref:Nak13,ref:Nak03}.
The exchange term of $V_C$ and the two-body term of $H_\mathrm{c.m.}$
are taken into account up to the pairing channel.

The numerical calculations have been implemented
with the Gaussian expansion method (GEM)~\cite{ref:GEM},
in which the single-particle (s.p.) or quasiparticle (q.p.) wave-functions
are expressed by a superposition of spherical Gaussian bases
having various ranges~\cite{ref:NS02,ref:Nak06,ref:Nak08}.
The basis functions are truncated by their orbital angular momentum $\ell$,
and we adopt $\ell\leq 7$ bases in this work.
For details of the basis functions, see Ref.~\cite{ref:Nak08}.
The QRPA calculations have been carried out
on top of the HFB solutions,
which are called HFB\,+\,QRPA calculations.
The RPA calculations take over the advantages of the GEM,
as examined in Ref.~\cite{ref:NMYM09},
handling loosely-bound s.p. states and spurious c.m. motion reasonably well
even with finite-range interactions.
As spurious modes associated with symmetry breaking
(\textit{i.e.}, Nambu-Goldstone mode) should have zero energy,
they give a measure of the accuracy of the numerical method.
In the QRPA, an additional spurious mode appears,
associated with the spontaneous breaking
of particle-number conservation in the HFB solution.
We have confirmed that this spurious mode is also separated well
in the spherical HFB\,+\,QRPA calculations with GEM,
whose energy $\omega_s$ satisfies $|\omega_s|^2 < 10^{-5}\,(\mathrm{MeV}^2$),
irrespective of nuclides and nucleonic interactions.
In the CHFB calculations in Sec.~\ref{sec:def},
the constraining term concerning the mass quadrupole moment
has been introduced into the MF framework~\cite{ref:SNM16,ref:MN18}.
The angular-momentum projection (AMP) is also applied to the CHFB solutions.
The AMP was implemented using the basis functions of the GEM
in Ref.~\cite{ref:AN22},
for expectation values of scalar operators including the Hamiltonian
on top of the axially-deformed Hartree-Fock (HF) solution.
The AMP with the GEM bases is here extended to transition matrix elements
between the eigenstates of the angular momentum~\cite{ref:RER02}
belonging to the same intrinsic state,
which is a solution of the axial HFB calculation.

\begin{figure}
\includegraphics[scale=0.75]{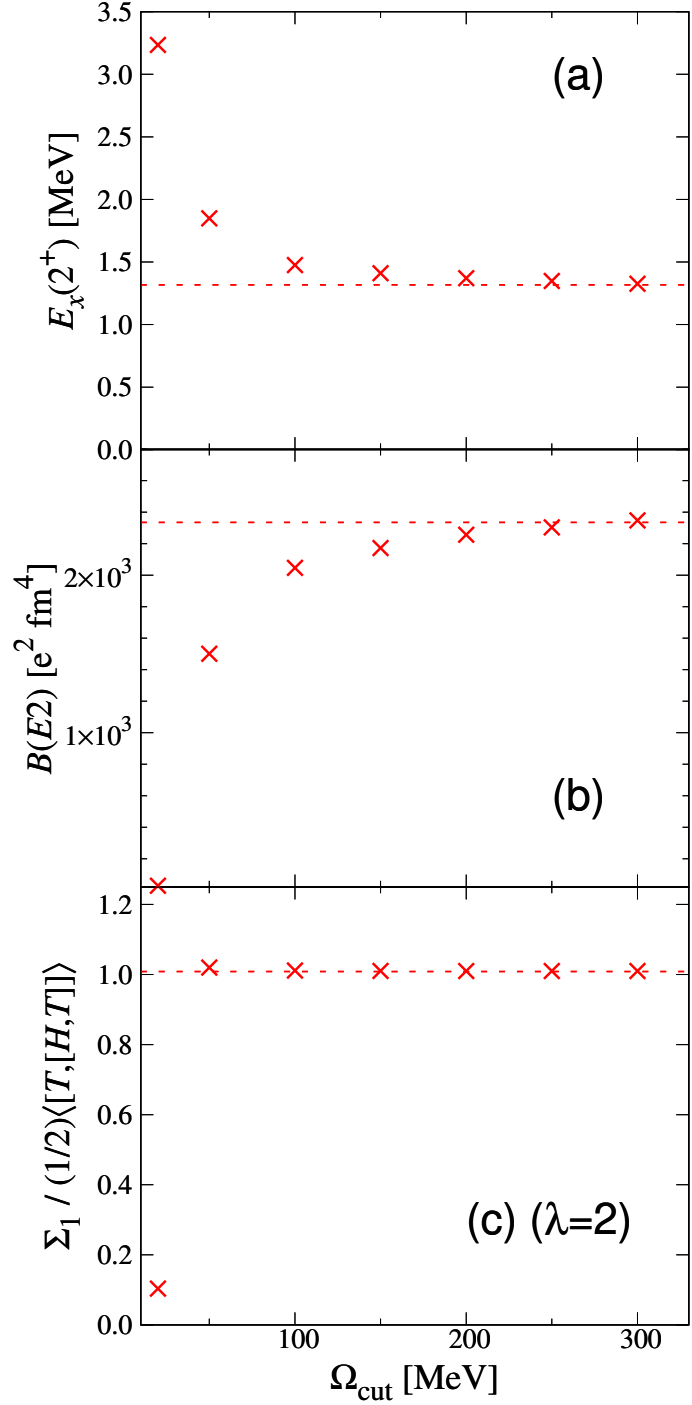}
\caption{Convergence of the QRPA results
  with respect to $\Omega_\mathrm{cut}$ at $^{116}$Sn,
  for (a) $E_x(2_1^+)$, (b) $B(E2;0^+_1\to 2^+_1)$,
  and (c) energy-weighted sum $\Sigma_1^{(\lambda=2)}$
  relative to $(1/2)\sum_\mu (-)^\mu \big\langle[T^{(\lambda=2,\mathrm{IS})}_\mu,
    [H,T^{(\lambda=2,\mathrm{IS})}_{-\mu}]]\big\rangle$.
  The dashed lines exhibit the values obtained with the full GEM bases.
\label{fig:conv}}
\end{figure}

This work includes the first application of the GEM and the M3Y-type interaction
to the HFB\,+\,QRPA calculations,
while applications to the HF\,+\,RPA approaches
were reported in Refs.~\cite{ref:Shi08,ref:NIS13,ref:IN15}.
The QRPA calculations need computation time and resources.
The excited states are often cut off via the sum of the q.p. energies
as $\varepsilon_{k_1}+\varepsilon_{k_2} < \Omega_\mathrm{cut}$,
where $k_1$ and $k_2$ stand for the q.p. states.
The convergence property with respect to $\Omega_\mathrm{cut}$
is shown in Fig.~\ref{fig:conv}.
In addition to $E_x(2^+_1)$ and $B(E2;0^+_1\to 2^+_1)$,
convergence for the energy-weighted sum of the isoscalar transition strength
is viewed,
\begin{equation}\begin{split}
  \Sigma_1^{(\lambda=2)}
  &= \sum_\nu E_x(2^+_\nu)\,\big|\langle 2^+_\nu||T^{(\lambda=2,\mathrm{IS})}
  ||0^+_1\rangle\big|^2\,;\\
  & T^{(\lambda=2,\mathrm{IS})}_{\mu} = \sum_{\tau=p,n} T^{(\lambda=2,\tau)}_\mu
  \,,~
  T^{(\lambda=2,\tau)}_\mu = \sum_{i\in\tau} r_i^2\,Y^{(2)}_\mu(\hat{\mathbf{r}}_i)
  \,,
\end{split}\label{eq:EWS}\end{equation}
which should be equal to the expectation value of the double commutator
at the ground state~\cite{ref:RS80,ref:NMYM09,ref:Nak17},
$(1/2)\sum_\mu (-)^\mu [T^{(\lambda=2,\mathrm{IS})}_\mu,
  [H,T^{(\lambda=2,\mathrm{IS})}_{-\mu}]]$.
We have confirmed that $E_x(2^+_1)$ and $B(E2;0^+_1\to 2^+_1)$
are convergent with $\Omega_\mathrm{cut}\approx 300\,\mathrm{MeV}$,
giving almost equal values to those calculated with all the GEM bases
used in the HFB calculation.
On the contrary, the convergence is not achieved
with $\Omega_\mathrm{cut}<100\,\mathrm{MeV}$.
Insufficient convergence may seriously influence
arguments in Secs.~\ref{sec:QRPA} and \ref{sec:def}.
It is noted that the energy-weighted sum converges
at a considerably lower $\Omega_\mathrm{cut}$,
and that good convergence for the energy-weighted sum
does not guarantee convergence for the low-energy collective state.
A similar convergence analysis was performed for Gogny interactions
in Refs.~\cite{ref:LDP15,ref:BG77}.
In Ref.~\cite{ref:LDP15},
the $E_x(2^+_1)$ and $B(E2;0^+_1\to 2^+_1)$ values
at $\Omega_\mathrm{cut}\approx 100\,\mathrm{MeV}$
are close to those with the full bases,
significantly faster than the present result in Fig.~\ref{fig:conv}.
This difference may be caused by the difference
in the employed bases and interaction.
We adopt $\Omega_\mathrm{cut}=300\,\mathrm{MeV}$
in the following QRPA calculations.

\section{Quadrupole collectivity under spherical symmetry via HFB\,+\,QRPA
  \label{sec:QRPA}}

As mentioned in Introduction,
the ground states of the even-$N$ Sn nuclei are conventionally considered
spherical.
It has seemed reasonable to
apply the HFB calculation to their ground states
and the HFB\,+\,QRPA framework to the first excited states,
keeping the spherical symmetry.

\begin{figure}
\includegraphics[scale=0.75]{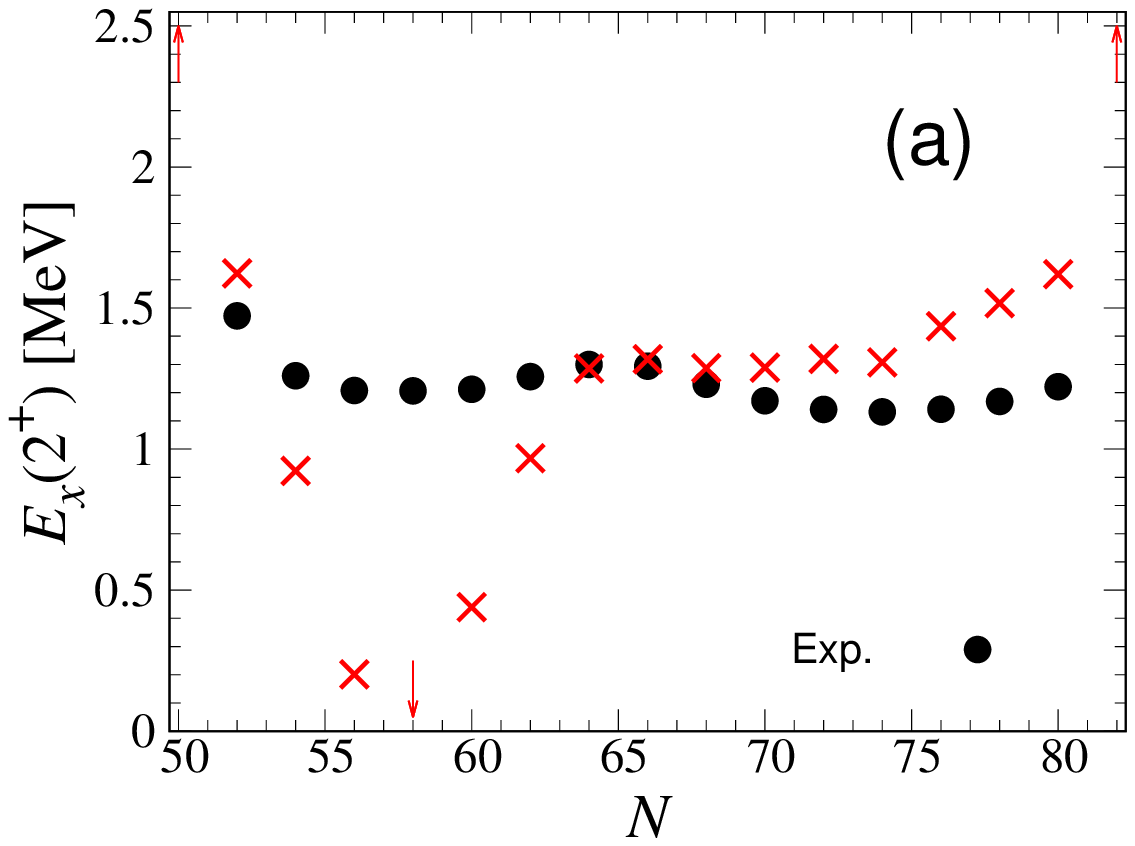}\\
\includegraphics[scale=0.75]{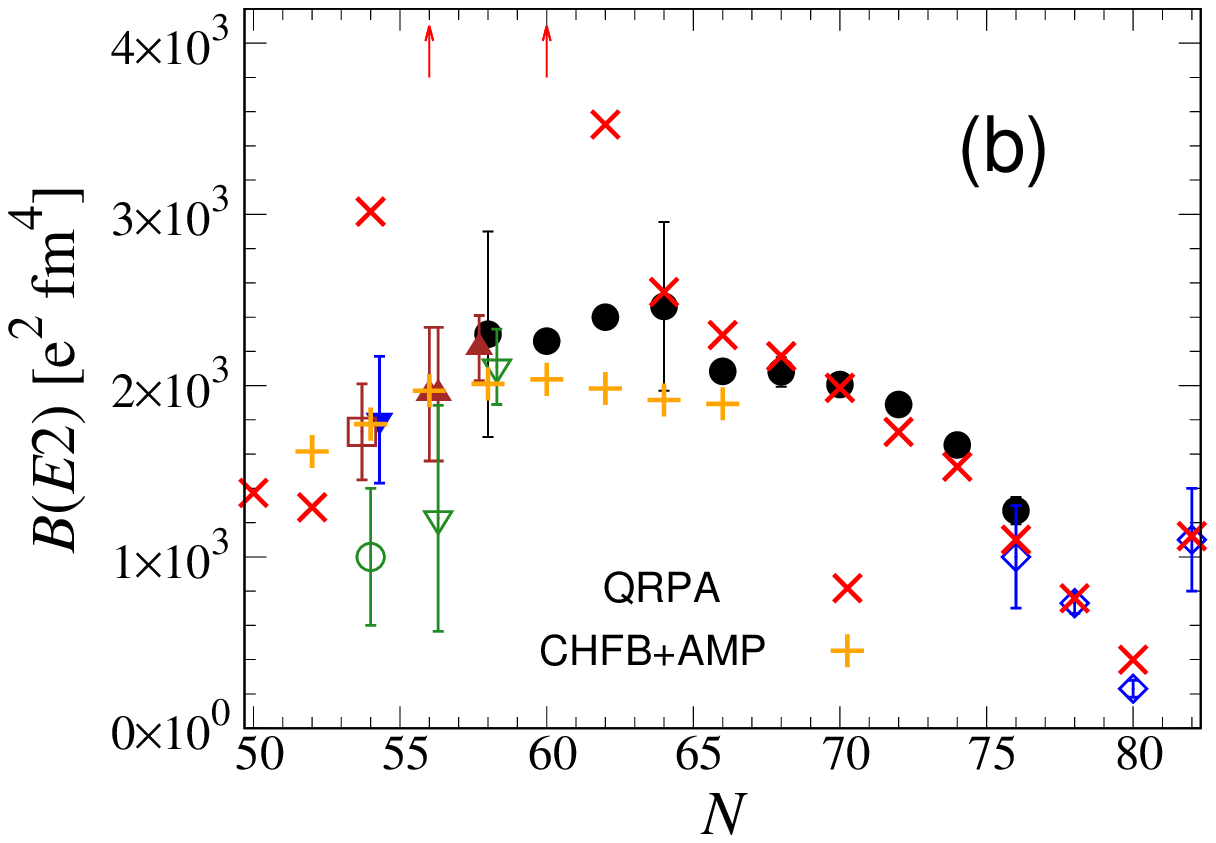}
\caption{(a) $E_x(2_1^+)$ and (b) $B(E2;0+_1\to 2^+_1)$ in $^{100-132}$Sn.
  The spherical HFB\,+\,QRPA results are presented by red crosses
  and compared with the experimental data,
  which are taken from Refs.~\cite{ref:NuDat} (black circles),
  \cite{ref:Rad05} (open blue diamonds),
  \cite{ref:Eks08} (brown triangles),
  \cite{ref:Gua13} (an open green circle),
  \cite{ref:Bad13} (a blue inverted triangle),
  \cite{ref:Doo14} (an open brown square),
  and \cite{ref:Sic20} (open green inverted triangles).
  Orange pluses for $^{102-116}$Sn are reference values
  obtained by the CHFB\,+\,AMP calculations,
  whose details are described in Sec.~\ref{subsec:AMP}.
\label{fig:QRPA}}
\end{figure}

In Fig.~\ref{fig:QRPA},
$E_x(2^+_1)$ and $B(E2;0^+_1\to 2^+_1)$ in $^{100-132}$Sn
obtained by the spherical HFB\,+\,QRPA calculations with M3Y-P6 are presented
in comparison with the experimental data~\cite{ref:NuDat,
  ref:Rad05,ref:Eks08,ref:Gua13,ref:Bad13,ref:Doo14}.
For the $E2$ transition operator,
the c.m. correction is taken into account
up to the one-body term~\cite{ref:EG2},
\begin{equation}
  T^{(E2)}_\mu = \sum_{\tau=p,n} \sum_{i\in\tau}
  e^{(E2)}_\tau\, T^{(\lambda=2,\tau)}_\mu\,;\quad
  e^{(E2)}_p = e\,\frac{(A-1)^2+(Z-1)}{A^2},~
  e^{(E2)}_n = e\,\frac{Z}{A^2}\,.
\end{equation}
The $E_x(2^+_1)$ values are out of the range of the figure at $^{100,132}$Sn.
The predicted value is $4.6\,\mathrm{MeV}$ at $^{100}$Sn,
and $4.7\,\mathrm{MeV}$ at $^{132}$Sn
compared to the measured value of $4.0\,\mathrm{MeV}$~\cite{ref:NuDat}.

In $64\leq N\leq 82$, both $E_x(2^+_1)$ and $B(E2;0^+_1\to 2^+_1)$
are in remarkable agreement with the experimental data.
It is emphasized that no adjustable parameters like the effective charge
are introduced in the present calculations.
These results support that the Sn nuclei have spherical shapes
at their ground states
and the quadrupole vibration constitutes the $2^+_1$ states,
in $64\leq N\leq 82$.
The HFB\,+\,QRPA calculations with the Gogny interaction
have also reproduced $E_x(2^+_1)$ and $B(E2;0^+_1\to 2^+_1)$
in $N\geq 66$~\cite{ref:PM14}
or in $70\lesssim N\leq 80$~\cite{ref:Nak11}.
Good agreement was recovered in $^{104,106}$Sn~\cite{ref:PM14,ref:Cor15}.
Difference between the results with the Gogny-D1S interaction
in Refs.~\cite{ref:PM14} and \cite{ref:Nak11}
may be attributed to the numerical methods
or the modification of the Hamiltonian for the QRPA calculation
in Ref.~\cite{ref:PM14}.
At $N=54$ and $62$, the present HFB\,+\,QRPA results deviate from the data,
but not very seriously.
However, the present HFB\,+\,QRPA results
have significant discrepancies with the data in $56\leq N\leq 60$.
At $N=58$, we have found an unstable QRPA solution
with an imaginary excitation energy.
It is noticed that the overestimate of $B(E2)$ in the neutron-deficient region
is opposite to the shell-model prediction within a single major shell.

In Ref.~\cite{ref:CTP12},
the spherical HFB\,+\,QRPA results with the Skyrme interactions were reported,
in which the SkM$^\ast$, SLy4 and SkX parameter-sets were employed.
Except for the SkM$^\ast$ interaction,
the measured $E_x(2^+_1)$ and $B(E2;0^+_1\to 2^+_1)$ have been reproduced
in $68\leq N\leq 80$.
With SLy4, quadrupole instability similar to the present one at $N=58$
takes place,
but in a wider and slightly shifted region.
The SkX interaction gives
relatively good $E_x(2^+_1)$ and $B(E2;0^+_1\to 2^+_1)$ values
even in the neutron-deficient region.
Because the SkX parameter-set was fitted to the experimental s.p. levels,
these results may imply important roles of the s.p. levels
in the neutron-deficient Sn nuclei,
to which we shall return in the subsequent section.

\begin{figure}
\includegraphics[scale=0.75]{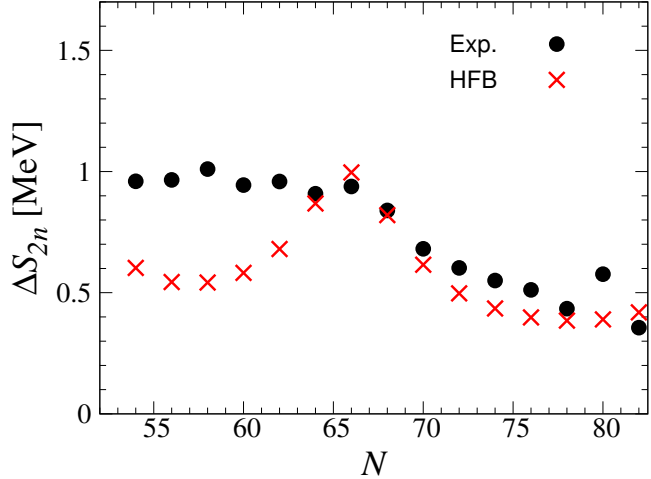}
\caption{$\mathit{\Delta}S_{2n}$ values
  from the spherical HFB calculations with M3Y-P6 (red crosses)
  and the experiments (black circles)~\cite{ref:AME20}.
\label{fig:dS2n}}
\end{figure}

As already mentioned,
the shell-model study in Ref.~\cite{ref:Tog18} suggested weak deformation
in the neutron-deficient region.
It was argued that the differential two-neutron separation energy,
\begin{equation}
  \mathit{\Delta}S_{2n}(Z,N) := S_{2n}(Z,N-2) - S_{2n}(Z,N)\,;\quad
  S_{2n}(Z,N) := E(Z,N-2) - E(Z,N)\,,
\end{equation}
would be an indicator of deformation.
We compare the spherical HFB results of $\mathit{\Delta}S_{2n}$
with the experimental values in Fig.~\ref{fig:dS2n}.
The calculations with M3Y-P6 are in good agreement with the experimental data
in $N\geq 64$,
supporting the sphericity in this region.
The enhancement of $\mathit{\Delta}S_{2n}$ around $N=66$
compared with the values in the larger-$N$ region is well reproduced;
therefore, it does not immediately imply deformation.
However, the spherical HFB approach fails to describe
$\mathit{\Delta}S_{2n}$ in $N\leq 62$.
This discrepancy may be attributed to deformation.
Whereas deformation influences nuclear radii~\cite{ref:BM1},
it has been shown that the measured charge radii
are reproduced by the spherical HFB approach
reasonably well~\cite{ref:Nak15,ref:Nak19} in $N\geq 58$,
suggesting that deformation is not strong even if it occurs.
It is desired to consider the possibility of weak quadrupole deformation
in this neutron-deficient region.

\begin{figure}
\includegraphics[scale=0.75]{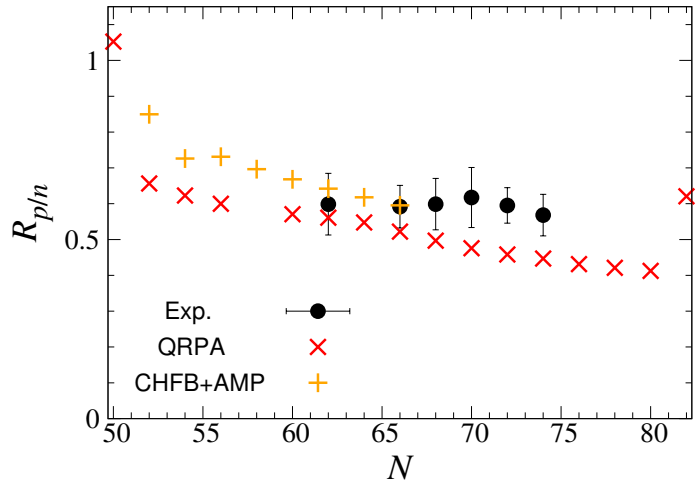}
\caption{Proton-to-neutron ratio of the quadrupole excitation
  $R_{p/n}$.
  The spherical HFB\,+\,QRPA results are presented by red crosses,
  and the reference values from CHFB\,+\,AMP by orange pluses.
  Experimental data are taken from Ref.~\cite{ref:Kun19}.
\label{fig:Mpn}}
\end{figure}

In the Sn nuclei,
low-energy excitation is expected to be dominated by neutrons.
However, the $E2$ strengths are carried by proton excitations
due to the core polarization.
It is interesting how many portions of excitation are attributed to protons.
Recently, ratios of the proton and neutron excitation strengths to $2^+_1$
have been reported experimentally~\cite{ref:Kun19}.
We here define the proton-to-neutron ratio,
\begin{equation}
  R_{p/n} = \frac{\langle 2^+_1||T^{(\lambda=2,\tau=p)}||0^+_1\rangle}
  {\langle 2^+_1||T^{(\lambda=2,\tau=n)}||0^+_1\rangle}\,,
\end{equation}
and compare the present HFB\,+\,QRPA results to the data in Fig.~\ref{fig:Mpn}.
Again they are in good agreement in the available region $62\leq N\leq 74$.
The agreement implies
that the core-polarization mechanism in the present HFB\,+\,QRPA framework
is appropriate.

\section{Deformability in neutron-deficient region\label{sec:def}}

As shown in Fig.~\ref{fig:QRPA},
too-low $E_x(2^+_1)$ and too-large $B(E2;0^+_1\to 2^+_1)$
are obtained by the spherical HFB\,+\,QRPA calculations in $^{104-112}$Sn,
indicating excessive quadrupole collectivity in the $54\leq N\leq 62$ region.
The too-strong quadrupole collectivity was also found
in the other self-consistent calculations
under the spherical symmetry~\cite{ref:CTP12,ref:PM14}.
The excessive collectivity holds for the SkX interaction,
in which the interaction parameters were adjusted
to the observed s.p. energies,
though less pronounced than the present results.
In the present calculation,
the overestimate of the collectivity is particularly serious
in $^{106-110}$Sn,
and the spherical HFB solution is even unstable
against the quadrupole deformation at $^{108}$Sn.
The too-large $B(E2)$ is a high contrast to the shell-model predictions
assuming the $^{100}$Sn inert core,
which underestimated the $B(E2)$ values in this region.
While the measured $E_x(2^+_1)$ values in $^{106-110}$Sn
are close to those in $^{112-130}$Sn,
the $B(E2)$ values are contradictory
among the spherical HFB\,+\,QRPA results,
the conventional shell-model results and the experimental data.
For this puzzle,
weak deformation was suggested in Ref.~\cite{ref:Tog18},
via the shell-model calculations allowing excitation out of the $^{100}$Sn core.
In this section,
we investigate whether quadrupole deformation could occur in these nuclei,
and whether the deformation d.o.f.s
could solve the controversy.

\subsection{Potential energy curves\label{subsec:PEC}}

\begin{figure}
\includegraphics[scale=0.75]{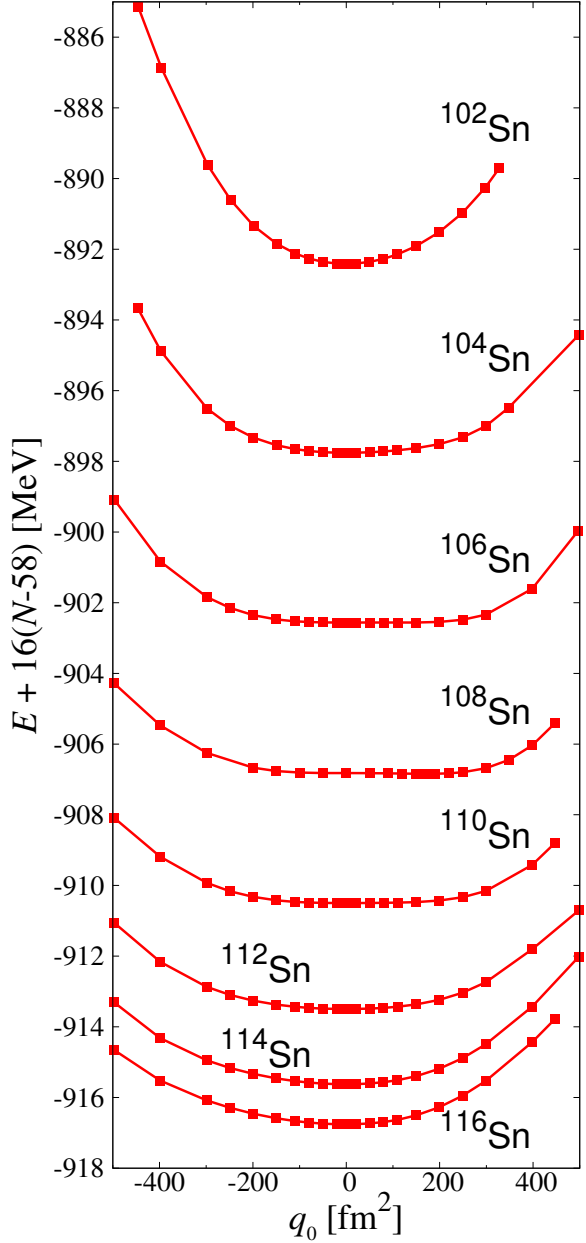}
\caption{Potential energy curves for $^{102-116}$Sn
  obtained from the axial CHFB calculations with M3Y-P6.
\label{fig:PEC_Sn102-116}}
\end{figure}

We implement the HFB calculations assuming the axial symmetry for $^{102-116}$Sn,
with constraining the mass quadrupole moment $q_0$~\cite{ref:SNM16},
where
\begin{equation}
  q_0 = \sqrt{\frac{16\pi}{5}}\,\Big\langle \sum_i (r_i-R)^2\,
  Y^{(2)}_0(\widehat{\mathbf{r}_i-\mathbf{R}})\Big\rangle\,;\quad
  \mathbf{R}=\frac{1}{A}\sum_i \mathbf{r}_i\,.
\label{eq:q0}\end{equation}
The potential energy curves (PECs) are displayed
in Fig.~\ref{fig:PEC_Sn102-116}.

In $^{102-104,112-116}$Sn,
the PECs have developed minimum at $q_0=0$.
In contrast,
the energies are almost flat against $q_0$ in $^{106-110}$Sn.
The range of $q_0$ of the flatness is $|q_0|\lesssim 200\,\mathrm{fm}^2$,
corresponding to the deformation parameter $|\beta_2|\lesssim 0.085$
if estimated via $q_0 = (3\beta_2/\sqrt{5\pi})AR^2$
with $R=1.12 A^{1/3}\,\mathrm{fm}$~\cite{ref:ZV17}.
These flat PECs suggest large shape fluctuation in these nuclei,
well accounting for the QRPA results in Fig.~\ref{fig:QRPA}
and indicating that the QRPA on top of the spherical HFB solution
is not appropriate to describe the structure of these nuclei.

The PECs in $^{106-110}$Sn rise at both ends of $q_0$.
Although we do not view the influence of triaxiality,
it is expected that the energies hardly depend on the degree of the triaxiality
at this weak deformation.
This flatness and the sudden rise of PECs remind us
of the $E(5)$ symmetry~\cite{ref:Iac00},
though the range of the deformation is not extensive
and we need the mass parameters for reliable evaluation
of the energy spectrum in the collective model~\cite{ref:WY_pc}.


\subsection{Balance between pairing and
  $n0g_{7/2}$-$n1d_{5/2}$ spacing\label{subsec:balance}}

The flatness of the PECs in  $^{106-110}$Sn in Fig.~\ref{fig:PEC_Sn102-116}
should be relevant to the quadrupole collectivity in these nuclei.
It is intriguing what makes the flat PECs.

The pairing should influence the PECs significantly,
having the tendency to keep the nuclear shape spherical.
To investigate the effects of the pairing,
we introduce a parameter $g$ that shifts the strength of the pairing channel,
\begin{equation}
  H_g = H + (g-1)\,V_N^\mathrm{pair}\,,
  \label{eq:gpair}\end{equation}
where $V_N^\mathrm{pair}$ is the pairing channel of $V_N$
in Eq.~\eqref{eq:Hamil}.
The HFB results in Fig.~\ref{fig:PEC_Sn102-116} are recovered with $g=1$,
and the pairing channel of $V_N$ is entirely removed with $g=0$.
We vary $g$ around $g=1$.

Since these nuclei lie around the middle of $N=50-64$,
in which neutrons mainly occupy the $0g_{7/2}$ and $1d_{5/2}$ orbits,
the difference of their s.p. energies should be relevant.
The roles of these two orbits in deformation
were already mentioned in Ref.~\cite{ref:Tog18}.
It is noted that the M3Y-P6 interaction gives
almost degenerate s.p. energies within the spherical HF framework;
$\epsilon(n0g_{7/2})-\epsilon(n1d_{5/2})=0.03\,\mathrm{MeV}$ at $^{110}$Sn,
which is compared to $1.4\,\mathrm{MeV}$
obtained with the Gogny-D1S interaction~\cite{ref:Nak11}.
Because of this near degeneracy,
these two orbits effectively form a large subshell
and could help to gain quadrupole collectivity,
contributing to the flat PEC,
as will be discussed below.
To investigate how the s.p. energy spacing influences,
we consider the following shift of the Hamiltonian,
\begin{equation}\begin{split}
  H_\xi &= H + \xi\cdot f_\mathrm{WS}(r)\cdot
  \big[-P_{n,d_{5/2}}+P_{n,g_{7/2}}\big]\,;\\
  &\qquad f_\mathrm{WS}(r) = \bigg[1+\exp\Big(\frac{r-R}{a}\Big)\bigg]^{-1}\,,
\end{split}\label{eq:dpot}\end{equation}
where $P_{\tau,\ell_j}$ stands for the projection operator
onto the s.p. levels having the orbital and summed angular momentum $(\ell j)$
for the particle type $\tau\,(=p,n)$.
The additional potential lowers (raises)
the neutron $d_{5/2}$ ($g_{7/2}$) levels,
and the parameter $\xi$ controls this size.
We take the parameters for the Woods-Saxon potential
as $R=r_0\,A^{1/3}$, $r_0=1.27\,[\mathrm{fm}]$
and $a=0.67\,[\mathrm{fm}]$~\cite{ref:BM1}.
Whereas not only $0g_{7/2}$ and $1d_{5/2}$ levels but also other nodal levels
feel the additional potential,
its main effects on the wave-function are for $0g_{7/2}$ and $1d_{5/2}$,
since the other orbitals are distant from the Fermi energy.

\begin{figure}
  \vspace*{-1cm}
\includegraphics[scale=0.75]{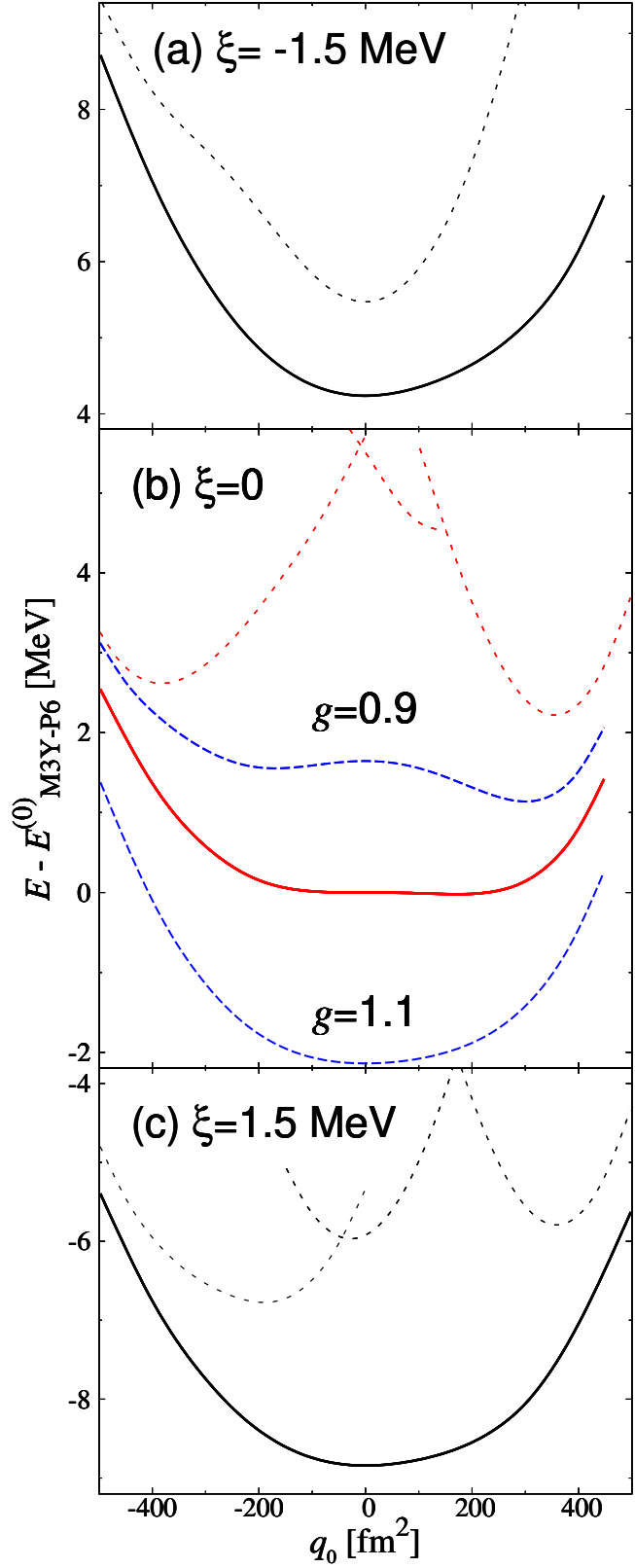}\vspace*{-0.5cm}
\caption{Potential energy curves for $^{108}$Sn with
  (a) $\xi=-1.5\,\mathrm{MeV}$, (b) $\xi=0$, and (c) $\xi=1.5\,\mathrm{MeV}$.
  The CHFB (CHF) results are displayed by the solid (short-dashed) lines.
  In (b), the results with $g=0.9$ and $1.1$ are depicted
  by the blue long-dashed lines.
  The energies are presented in terms of their difference
  from the spherical HFB energy with M3Y-P6 (viz., $\xi=0$, $g=1$).
\label{fig:PEC_Sn108_detail}}
\end{figure}

The $g$- and $\xi$-dependence of the PEC is depicted for $^{108}$Sn
in Fig.~\ref{fig:PEC_Sn108_detail}.
The results are similar for $^{106,110}$Sn.
From the $g$-dependence shown in Fig.~\ref{fig:PEC_Sn108_detail}\,(b),
we confirm that the spherical shape becomes the more favored
for the stronger pairing,
and the PEC significantly depends on $g$.
We have a well-developed minimum at $q_0=0$ with $g=1.1$,
while the $q_0=0$ state is unstable with $g=0.9$.
Thus the pairing is one of the crucial ingredients
for the flat PEC with the M3Y-P6 (viz., $g=1$) interaction.
It is recalled that pairing properties with M3Y-P6 have been examined
for a large number of semi-magic nuclei~\cite{ref:Nak13}.

Concerning the $\xi$-dependence,
let us first look at the constrained-HF (CHF) results,
which are presented by the short-dashed lines
in Fig.~\ref{fig:PEC_Sn108_detail}.
With $\xi=-1.5\,\mathrm{MeV}$,
the spherical (viz., $q_0=0$) state gives a well-developed minimum,
in which $n0g_{7/2}$ is fully occupied and $n1d_{5/2}$ is empty.
In contrast,
we find deformed minima, a prolate and an oblate ones, with $\xi=0$.
No spherical configurations form a well-developed minimum.
This is an effect of the near degeneracy of $n0g_{7/2}$ and $n1d_{5/2}$,
which makes neutrons occupy $0g_{7/2}$ and $1d_{5/2}$
with almost equal probabilities.
Then both the $n0g_{7/2}$ and $n1d_{5/2}$ orbits are halfly occupied
at $N\approx 58$.
Since a half of the magnetic substates favor the prolate shape
and the other half the oblate shape,
energy minima are developed on both sides.
In practice,
$|m|=1/2$ and $3/2$ levels primarily composed of $g_{7/2}$ and $d_{5/2}$
are occupied at the prolate minimum,
while admixture of $d_{3/2}$ and $s_{1/2}$ components are not negligible
at the oblate minimum.
With $\xi=1.5\,\mathrm{MeV}$, a spherical minimum emerges again
at which $n1d_{5/2}$ is mostly occupied and $n0g_{7/2}$ is partly populated,
as well as deformed minima.
When the pairing is set on,
a spherical configuration gains a significant amount of the pair correlation.
Thus the flat PEC for $^{106-110}$Sn in Figs.~\ref{fig:PEC_Sn102-116}
and \ref{fig:PEC_Sn108_detail}\,(b)
is connected to the developed prolate and oblate minima
and the absence of the spherical minimum at the CHF level,
which takes place because $n0g_{7/2}$ and $n1d_{5/2}$ merge
and behave like a single large orbital.

As argued above, it seems reasonable to consider
that the deformability in the neutron-deficient Sn nuclei
is triggered by the neutrons occupying $0g_{7/2}$ and $1d_{5/2}$.
In Refs.~\cite{ref:Tog18,ref:KSMS21,ref:Bad13},
a role of $p0g_{9/2}$ was stressed through the studies
within a restricted model space assuming effective charges.
The present results on the role of the neutrons are not in contradiction to it,
as the protons occupying $0g_{9/2}$ are more or less excited
by the deformation.
It should still be emphasized that the enhancement of the proton collectivity
is subsidiary, induced by the neutron collectivity.

\subsection{Possibility of sphericity revisited\label{subsec:sphericity}}

In the previous subsection,
we have observed that the balance between the pairing effects
and those of the near degeneracy of $n0g_{7/2}$-$n1d_{5/2}$ is crucial
for the quadrupole collectivity in the neutron-deficient Sn nuclei.
While it has been confirmed that
the present effective Hamiltonian successfully describes
the ground-state properties of nuclei in a wide mass range~\cite{ref:Nak20},
there might be a room to ask if it is sufficiently precise
to describe the pairing and shell structure in this particular region.
In this subsection,
we return to the spherical HFB and HFB\,+\,QRPA calculations,
and investigate whether the measured $E_x(2^+_1)$ and $B(E2)$
can be accounted for within the spherical picture
if we allow slight variation of the pairing or the s.p. energies.

\begin{figure}
  \vspace*{-1cm}
\includegraphics[scale=0.65]{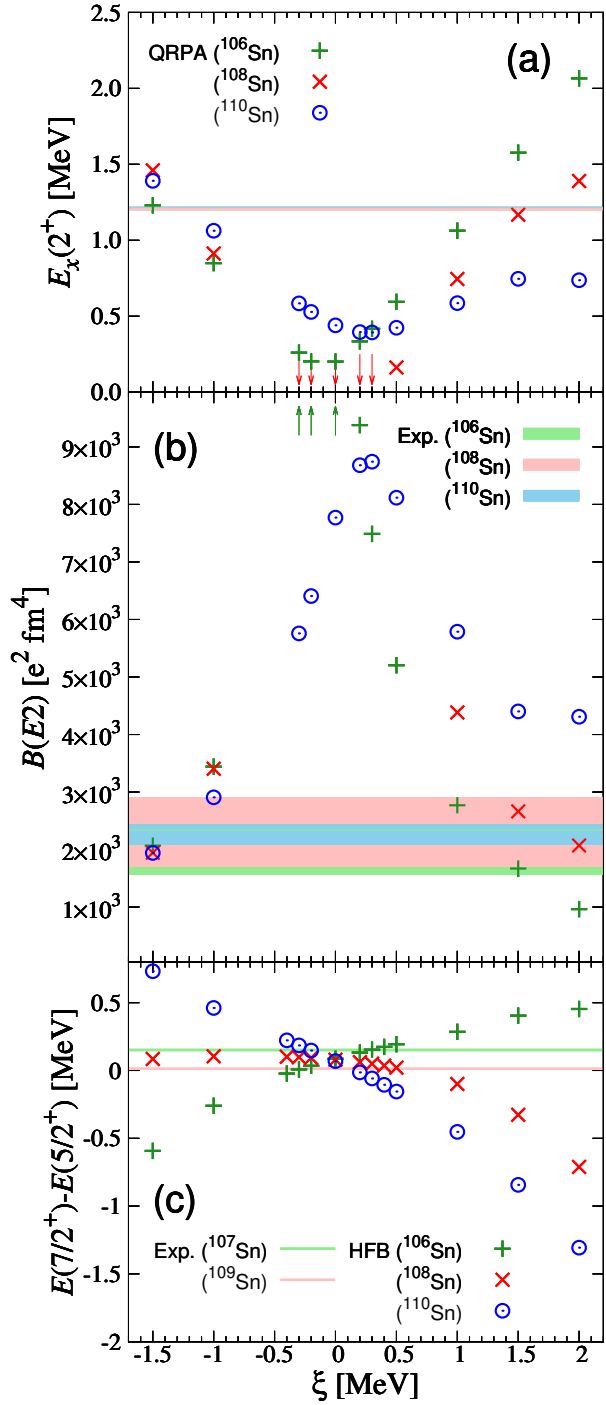}\vspace*{-0.5cm}
\caption{$\xi$-dependence of (a) $E_x(2^+_1)$, (b) $B(E2;0^+_1\to 2^+_1)$,
  and (c) $E(7/2^+)-E(5/2^+)$.
  The spherical HFB or HFB\,+\,QRPA results are depicted
  by green pluses ($^{106}$Sn), red crosses ($^{108}$Sn)
  and open blue circles ($^{110}$Sn).
  The arrows in (b) indicate that the HFB\,+\,QRPA results
  are out of the range of this plot.
  In $^{108}$Sn, the QRPA calculation gives an unstable solution
  in $|\xi|\leq 0.3$.
  In (a) and (b),
  the experimental values are displayed by the light-green ($^{106}$Sn),
  pink ($^{108}$Sn) and skyblue ($^{110}$Sn) bands,
  which represent the range of the errors,
  although the measured $E_x(2^+_1)$ values are so close among $^{106,108,110}$Sn
  that they are almost indistinguishable.
  In (c), the experimental values are shown
  by the light green ($^{107}$Sn) and pink ($^{109}$Sn) lines.
\label{fig:sph-dpot}}
\end{figure}

As the s.p. energy difference is significant,
energies of lowest-lying states in odd-$N$ Sn nuclei
should be investigated simultaneously.
Experimentally, $5/2^+$ and $7/2^+$ levels lie very closely
at $^{105,107,109}$Sn.
These levels are assigned to the $d_{5/2}$ and $g_{7/2}$ q.p. levels.
In Fig.~\ref{fig:sph-dpot},
$\xi$-dependence of the q.p. energy differences
$\varepsilon(n0g_{7/2})-\varepsilon(n1d_{5/2})$ in the spherical HFB calculations
is shown
in comparison with the measured energy differences between $5/2^+$ and $7/2^+$,
as well as of $E_x(2^+_1)$ and $B(E2;0^+_1\to 2^+_1)$ at $^{106,108,110}$Sn.
Although $|\xi|>1\,\mathrm{MeV}$ is needed to reproduce
$E_x(2^+_1)$ and $B(E2;0^+_1\to 2^+_1)$,
the observed close energy between $5/2^+$ and $7/2^+$ at $^{107,109}$Sn
can be obtained only with $|\xi|\lesssim 0.5\,\mathrm{MeV}$.

At a glance,
one might think that $\xi\approx -1\,\mathrm{MeV}$ gives acceptable results,
not quite contradictory to the measurements.
However, this shift of the s.p. energies indicates
that $n0g_{7/2}$ should lie substantially lower than $n1d_{5/2}$,
which is reflected in the lowest levels of odd-$N$ Sn nuclei
in $51\leq N\lesssim 58$.
Experimentally, it is suggested~\cite{ref:NuDat}
that the ground states of $^{101-109}$Sn have $5/2^+$.
This property cannot be reproduced with $\xi\approx -1\,\mathrm{MeV}$.
The s.p. energy shift as simulated by $\xi\approx -1\,\mathrm{MeV}$
is not likely.

It is commented that,
despite a relatively good agreement with the data
on $E_x(2^+_1)$ and $B(E2;0^+_1\to 2^+_1)$
within the HFB\,+\,QRPA~\cite{ref:PM14},
the spherical HFB calculations with D1S do not describe
$\varepsilon(n0g_{7/2})-\varepsilon(n1d_{5/2})$ well,
because these two s.p. levels keep distant~\cite{ref:Nak11}.
The difference in the q.p. energies is $0.32$ to $0.42\,\mathrm{MeV}$
in $^{106-110}$Sn.

\begin{figure}
\includegraphics[scale=0.7]{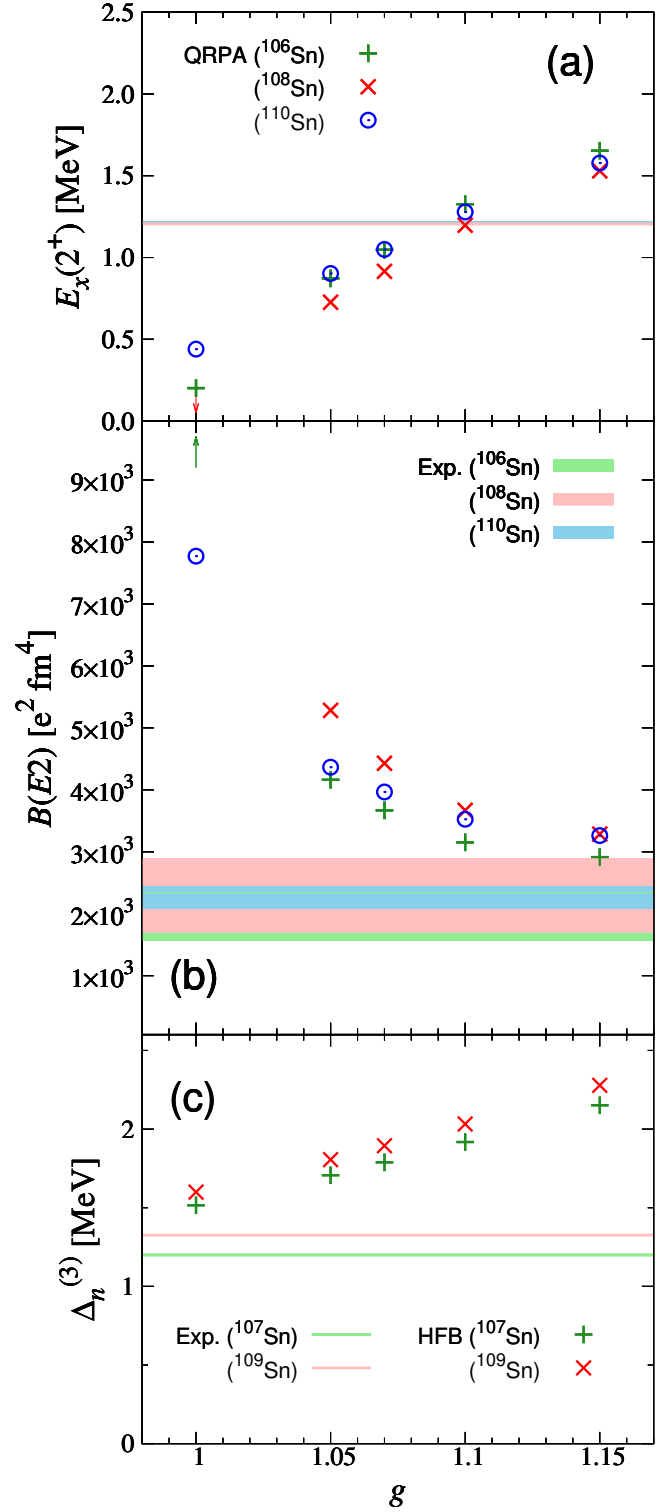}
\caption{$g$-dependence of (a) $E_x(2^+_1)$, (b) $B(E2;0^+_1\to 2^+_1)$,
  and (c) $\Delta_n^{(3)}$.
  For conventions of (a) and (b), see Fig.~\protect\ref{fig:sph-dpot}.
  In (c), the spherical HFB results are depicted
  by the green pluses ($^{107}$Sn) and the red crosses ($^{109}$Sn),
  while the experimental values are displayed
  by the light-green ($^{107}$Sn) and the pink ($^{109}$Sn) lines.
\label{fig:sph-cpair}}
\end{figure}

In Fig.~\ref{fig:sph-cpair},
$g$-dependence of the even-odd energy difference for $^{107,109}$Sn,
where
\begin{equation}
  \Delta_n^{(3)}(Z,N) := \frac{(-)^N}{2}
    \big[E(Z,N+1)-2E(Z,N)+E(Z,N-1)\big]\,,
    \label{eq:D-3p}\end{equation}
is shown,
as well as of $E_x(2^+_1)$ and $B(E2;0^+_1\to 2^+_1)$.
The energies of $N=\mathrm{odd}$ nuclei have been calculated
within the spherical HFB in the equal-filling approximation~\cite{ref:EFA}.
Whereas $g>1$ is needed to reproduce $E_x(2^+_1)$ and $B(E2;0^+_1\to 2^+_1)$,
it is contradictory to $\Delta_n^{(3)}$.
Moreover, we have found that it is difficult to reproduce
both $E_x(2^+_1)$ and $B(E2;0^+_1\to 2^+_1)$ simultaneously,
irrespective of $^{106,108,110}$Sn;
while $g\approx 1.1$ is good for the former,
the latter indicates $g\geq 1.15$.

From these results with shifting the s.p. energies and the pairing,
we conclude it difficult to consider the neutron-deficient Sn nuclei
keeping the spherical shape.

\subsection{$B(E2)$ via angular momentum projection
  \label{subsec:AMP}}

As argued in this section so far,
it seems reasonable to consider that weak quadrupole deformation
including fluctuation
plays significant roles in $E_x(2^+_1)$ and $B(E2;0^+_1\to 2^+_1)$
of the neutron-deficient Sn nuclei, particularly $^{106-110}$Sn.
For $^{108}$Sn, we find the absolute minimum at $q_0=170\,\mathrm{fm}^2$,
to which the protons contribute by $39\,\%$.
However, this deformation is too weak
to apply the formula for well-deformed nuclei~\cite{ref:BM2}.
Indeed, while it is close to the edge of the flat PEC,
this minimum at $^{108}$Sn gives
$B(E2;0^+_1\to 2^+_1)\approx 435\,e^2\mathrm{fm}^4$
if estimated via the formula,
\begin{equation}
  B(E2;0^+_1\to 2^+_1)\approx e^2\,(0\,0\,2\,0|2\,0)^2\,
  \Big\langle \sum_{i\in p} (r_i-R)^2\,
  Y^{(2)}_0(\widehat{\mathbf{r}_i-\mathbf{R}})\Big\rangle^2\,,
\label{eq:BE2_def}\end{equation}
far smaller than the measured value.
We further discuss whether weak quadrupole deformation
can account for the measured $E2$ strengths in the neutron-deficient Sn nuclei.

The flat PECs in $^{106-110}$Sn suggest
that admixture of intrinsic states with various deformations occurs easily.
It is desirable to superpose MF wave-functions that have different $q_0$'s.
One of the approaches in this line is mapping the constrained MF results
to the collective model, \textit{e.g.}, the Bohr model~\cite{ref:BM2}.
The 5DCH was constructed from the CHFB results with the Gogny-D1S interaction
and applied to the Sn nuclei in Ref.~\cite{ref:PM14}.
However, the 5DCH results are not better for $B(E2)$
than the HFB\,+\,QRPA results.
It is also commented that there remain problems
in evaluating the mass parameters
when constructing the Hamiltonian
consisting of the collective d.o.f.s~\cite{ref:WHN21}.
The generator-coordinate method (GCM) is another approach,
in which MF wave-functions along collective coordinates
are superposed explicitly.
However, no computer codes for the GCM calculation have yet been available
with the M3Y-type semi-realistic interactions.
We also note a problem in handling the density-dependent interaction
within the GCM~\cite{ref:Rob10}.
In applying an effective Hamiltonian developed for the MF calculations
to beyond-MF studies,
we might need to readjust the interaction parameters,
or subtract correlation effects already contained,
as has been argued for the second RPA~\cite{ref:Tse07,ref:GGE15}.
Instead of superposing the MF solutions,
we take a single MF solution,
which may approximately represent physical quantities for surrounding solutions.
As an appropriate superposition of the MF states could be essential
to reproduce the energies,
we shall focus on the $E2$ transition strengths.
Rather than precise evaluation,
we attempt to obtain a helpful reference
to argue whether weak deformation including fluctuation
is likely in the neutron-deficient Sn nuclei.

\begin{figure}
\includegraphics[scale=0.75]{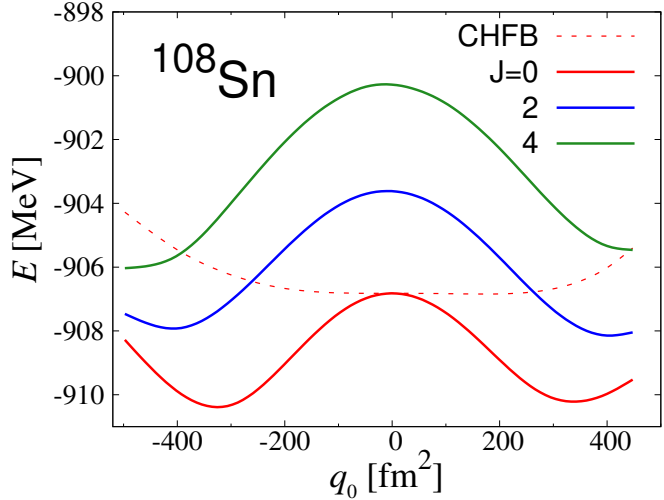}
\caption{$q_0$--dependence of energies of $0^+$, $2^+$ and $4^+$ states
  for $^{108}$Sn by the CHFB\,+\,AMP calculations with M3Y-P6.
\label{fig:108Sn-AMP}}
\end{figure}

We apply the AMP for this purpose.
The AMP can give a reasonable estimate of $B(E2)$
even for weakly-deformed intrinsic states~\cite{ref:Abe23},
because it explicitly handles the wave-functions.
We carry out the AMP on top of the CHFB solutions.
Energies of the eigenstates of angular momentum $J$ are given
as a function of $q_0$,
as exemplified for $^{108}$Sn in Fig.~\ref{fig:108Sn-AMP}.
As a reference for $B(E2;0^+_1\to 2^+_1)$,
we use the value at $q_0$ corresponding to the energy minimum
of the $J=0$ component, 
which is estimated by interpolation.
These reference values were plotted for $^{102-116}$Sn
in Fig.~\ref{fig:QRPA}\,(b) by orange pluses.
They are close to the measured values in $^{104-110}$Sn.
Although they are only the reference values
and should not be taken too seriously,
they demonstrate that weak deformation can account for
the $B(E2)$'s in the neutron-deficient Sn nuclei.
We note $\big[\min_{q_0}E(2^+)-\min_{q_0}E(0^+)\big]\gtrsim 2\,\mathrm{MeV}$,
the energy difference between the minima of individual $J$,
which is higher than the observed $E_x(2^+_1)$.
The superposition of the MF states is inevitable
for a complete understanding of the low-energy quadrupole collectivity
of the neutron-deficient Sn nuclei,
which is left for future work.

The proton-to-neutron ratios of the transition matrix elements $R_{p/n}$
obtained by the CHFB\,+\,AMP calculations
were plotted in Fig.~\ref{fig:Mpn}.
The ratios are not so different from those in the spherical HFB\,+\,QRPA.
Recall that, in the HFB\,+\,QRPA results,
the protons fully occupy all the orbits up to $0g_{9/2}$ at the ground states,
and the $E2$ transition occurs via the core polarization.
Furthermore, the ratios are insensitive to $q_0$
when we apply the AMP to each CHFB state.
This observation supports the interpretation
that the neutrons trigger the deformation, dragging protons.
As discussed in Sec.~\ref{subsec:balance},
the near degeneracy of $n0g_{7/2}$ and $1d_{5/2}$ plays an essential role
in the deformation.

\section{Summary\label{sec:summary}}

We have investigated quadrupole collectivity of the lowest-lying states
of the $N=50-82$ Sn nuclei,
focusing on $E_x(2^+_1)$ and $B(E2;0^+_1\to 2^+_1)$,
by applying the self-consistent approaches
with the semi-realistic interaction M3Y-P6.
This nucleonic interaction has been fixed
so as to describe the nuclear properties globally,
and the calculations contain no adjustable parameters.
Both $E_x(2^+_1)$ and $B(E2;0^+_1\to 2^+_1)$ are well reproduced
by the spherical HFB\,+\,QRPA calculations in $N\geq 64$,
endorsing the sphericity of these nuclei.
In $54\leq N\leq 62$,
the spherical HFB\,+\,QRPA calculations overestimate $B(E2)$,
oppositely to the shell-model predictions within the one major shell.

Via the constrained HFB calculations,
we have found that the neutron-deficient Sn nuclei are soft
against the quadrupole deformation.
In particular, the potential energy curves (PECs) are almost flat
in the range of $|q_0|\lesssim 200\,\mathrm{fm}^2$ in $^{106-110}$Sn.
This property limits the applicability of the HFB\,+\,QRPA approach.
In the neutron-deficient Sn nuclei,
the $n0g_{7/2}$ and $n1d_{5/2}$ orbits lie very close in energy.
This near degeneracy enhances the quadrupole deformation,
and produces the flat PEC in balance with the pairing
that favors sphericity.
This picture has been confirmed by the calculations
with shifting the single-particle energies
or varying the pairing strength,
and with the angular-momentum projection
on top of the constrained HFB solutions.
All the calculations keeping the sphericity are contradictory
to relevant experimental data,
and weak deformation with fluctuation is likely.
We have also found that the proton-to-neutron ratios
of the transition matrix elements
are insensitive to the neutron number and the deformation,
consistent with the interpretation
that the deformation is triggered by neutrons.

The self-consistent approaches, \textit{e.g.}, the HFB\,+\,QRPA,
enable us to describe the quadrupole collectivity of nuclei
without adjustable parameters.
However, one should be careful in the effective interaction.
It has been found that the shell structure and the pairing are key
to precisely describing the low-energy quadrupole collectivity
of the neutron-deficient Sn nuclei.
As appropriately describing the shell structure and the pairing properties,
the present semi-realistic interaction seems suitable
for investigating the low-energy quadrupole collectivity of these nuclei.
Indeed, it has supplied a convincing picture of the structure of the Sn nuclei
from the proton-rich to neutron-rich sides.
Note again that this interaction is
almost free from the unphysical instabilities against excitations,
as examined in the homogeneous nuclear matter.
It is interesting to apply the interaction to extensive studies
including beyond-MF calculations,
which may give a quantitative description
of many nuclei including the neutron-deficient Sn.

\begin{acknowledgments}
  The authors are grateful to N.~Shimizu and K.~Washiyama
  for the discussions.
  The numerical calculations were performed by Oakforest PACS and Wisteria/BDEC,
  whose resources were provided
  by the Multidisciplinary Cooperative Research Program
  in Center for Computational Sciences, University of Tsukuba,
  Yukawa-21 at Yukawa Institute of Theoretical Physics, Kyoto University,
  and HITAC SR24000 at the Institute of Management and Information Technologies,
  Chiba University.
\end{acknowledgments}


\end{document}